\documentclass[12pt,aps,prb,preprint]{revtex4}   

\usepackage{amsmath}    
\usepackage{amssymb}
\usepackage{graphicx}   

\usepackage[latin1]{inputenc}     
\usepackage[T1]{fontenc}

\usepackage{float}   

\draft

\begin{document}

\title{A complete analysis of the Stern-Gerlach experiment using Pauli spinors}
\author{Michel Gondran}
 \affiliation{EDF, Research and Development, 92140 Clamart.}
 \email{michel.gondran@chello.fr}   
\author{Alexandre Gondran}
 \affiliation{SeT, Université des Technologies de Belford Montbéliard}
 \email{alexandre.gondran@utbm.fr}   

\begin{abstract}

The Stern-Gerlach experiment is the fundamental experiment in
order to exhibit the quantization of spin and understand the
measurement problem in quantum mechanics. However, although the
Stern-Gerlach experiment plays an essential role in the teaching
of quantum mechanics, no complete analysis of this experiment
using Pauli spinors is presented in the pedagogical literature.
This paper presents such an analysis and develops implications for
the theory of quantum measurement.

We first propose an analytic expression of both the wave function
and the probability density in the Stern-Gerlach experiment. Our
explicit solution is obtained via a complete integration of the
Pauli equation over time and space. The probability density
evolution describes a slipping of the wave packet into two
separate packets due to the measurement device, but it cannot
account for impacts.

We therefore calculate the de Broglie-Bohm trajectories, which not
only explain impacts naturally, but also accounts for the spin
quantization following the magnetic field gradient. It is then
possible to propose a clear explanation of measurement effects in
the Stern-Gerlach experiment.

\end{abstract}

\maketitle

\section{Introduction}
\begin{figure}[H]
\begin{center}
\includegraphics[width=0.6\linewidth]{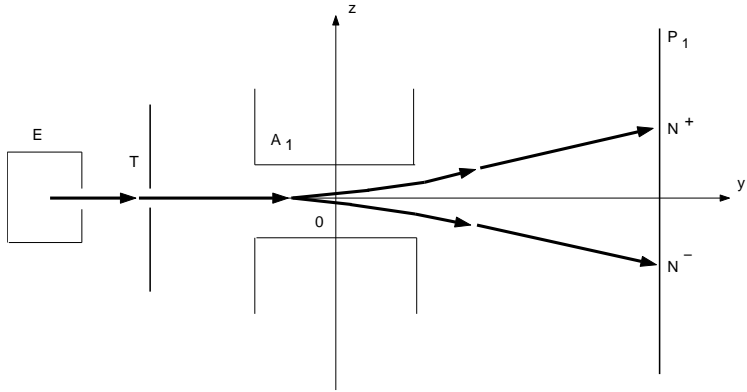}
\caption{\label{fig:schema-SetG}Schematic configuration of the
Stern-Gerlach experiment.}
\end{center}
\end{figure}

As they were studying the deviation of a silver atoms beam in a
highly inhomogeneous magnetic field (cf.
FIG.~\ref{fig:schema-SetG}) Stern and Gerlach
(1922)\cite{SternGerlach} found empirical results which challenged
common sense prediction. Instead of being scattered, the beam
split into two symetric beams, which produced two spots of equal
intensity on a screen, at equal distances from the axis of the
original beam.

This experiment motivated the introduction of spin quantization as
an intrinsic magnetic moment. It also clearly exhibits the
measurement problem in quantum mechanics, which remains an active
field of study.\cite{Bohr,Zurek,Schlosshauer}

This paper brings new elements, such as the analytic expression of
the wave function and the probability density in the Stern-Gerlach
experiment. The explicit solution is obtained via a complete
integration of the Pauli equation over time and space. As far as
we know, the analytic presentation of the Stern-Gerlach experiment
in text-books\cite{FeynmanCours} is only given in time, not in
space. The first explicit calculation in space and time of
Stern-Gerlach experiment was given by Dewdney \textit{et
al}\cite{Dewdney_1986}, but it remains incomplete, as is also
incomplete their explicit solution in space and time of the Dirac
equation.\cite{Challinor} Recent presentations in space and time
are only given with numerical simulations.\cite{Frana_1992}

The analytic solution presented here gives an explicit expression
of the probability density's evolution in space, explaining the
semi-classical presentation and showing the wave function
separation. However, this continuous evolution in space of the
wave packet into two wave packets does not account for the
particle impacts. We therefore also calculate the de Broglie-Bohm
trajectories \cite{deBroglie_1951} as we formerly did in the case
of Young's double slit experiment.\cite{Gondran_2005} These
trajectories not only provide a natural explanation for the impact
of particles, but also describe the spin quantization along the
z-axis. It is then possible to propose a clear explanation of
measurement in the Stern-Gerlach experiment.

The explicit solution in terms of Pauli spinors and the evolution
of the probability density for the Stern-Gerlach experiment are
presented in section 2. The de Broglie-Bohm  trajectories, as
defined by the interpretation of the Pauli equation by
Bohm\cite{Bohm_1955} and Takabayasi,\cite{Takabayasi_1954} are
simulated in section 3. Details of the calculations are provided
in Appendix A.


\section{The probability density calculation in the Stern-Gerlach
experiment}

Silver atoms contained in the oven E (Fig.~\ref{fig:schema-SetG})
are heated to a high temperature and escape through a narrow
opening. A second aperture, T, selects those atoms whose velocity,
$\textbf{v}_0$, is parallel to the y-axis. The atomic beam crosses
the gap of the electromagnet $A_{1}$ before condensing on the
screen, $P_{1}$ . The magnetic moments of the silver atoms before
crossing the electromagnet are oriented randomly (isotropically).
In the beam, we represent the atoms by their wave function ; one
can suppose that at the entrance to the electromagnet, $A_{1}$,
and at the initial time $t=0$, each atom prepared can be described
by a Gaussian spinor in x and z:
\begin{equation}\label{eq:psi-0}
    \Psi^{0}(x,z) = (2\pi\sigma_{0}^{2})^{-\frac{1}{2}}
                      e^{-\frac{(z^2+x^2)}{4\sigma_0^2}}
                      \binom{\cos \frac{\theta_0}{2}e^{ i\frac{\varphi_0}{2}}}
                            {i\sin\frac{\theta_0}{2}e^{-i\frac{\varphi_0}{2}}}.
\end{equation}

The variable y will be treated in a classical way with $y= vt$.
For a silver atom, one has $m = 1.8\times 10^{-25}$ kg, $v_0 =
500$\ m/s(with T=1000°K), $\sigma_0$=10$^{-4}$m.

In (\ref{eq:psi-0}), $\theta_0$ and $\varphi_{0}$ are the polar
angles characterizing the initial orientation of the magnetic
moment, $\theta_0$ corresponds to the angle with the z-axis. This
initial orientation being randomized, one can suppose that
$\theta_0$ is drawn in a uniform way from $[0,\pi]$ and that
$\varphi_0$ is drawn in a uniform way from $[0,2\pi]$. In this
way, we give a model of a mixture of pure states.

The evolution of the spinor $\Psi=\binom{\psi_{+}}
                            {\psi_{-}}$ in a magnetic field $\textbf{B}$ is then given by the Pauli
                            equation \cite{Platt}:

\begin{equation}\label{eq:Pauli}
    i\hbar \left( \begin{array}{c} \frac{\partial \psi _{+}}{\partial t}
                                   \\
                                   \frac{\partial \psi _{-}}{\partial t}
                  \end{array}
           \right)
    =-\frac{\hbar ^{2}}{2m} \nabla^2
                           \left( \begin{array}{c} \psi _{+}
                                                   \\
                                                   \psi _{-}
                                  \end{array}
                           \right)
     +\mu _{B}\textbf{B}\boldsymbol\sigma \left( \begin{array}{c} \psi _{+}
                                                                  \\
                                                                  \psi _{-}
                                                 \end{array}
                                          \right)
\end{equation}
where $\mu_B=\frac{e\hbar}{2m_e}$ is the Bohr magneton and where
$\boldsymbol\sigma=(\sigma_{x},\sigma_{y},\sigma_{z})$ corresponds
to the three Pauli matrices.

The particle first enters an electromagnetic field $\textbf{B}$
directed along the z-axis, $B_{x}=B'_{0} x$, $B_{y}=0$,
$B_{z}=B_{0} -B'_{0} z$, with $B_{0}=5$ Tesla and $B'_{0}=\left|
\frac{\partial B}{\partial z}\right| = 10^3$ Tesla/m over a length
$\Delta l=1~cm$. Such a vector $\textbf{B}$ satisfies Maxwell's
equations, since $div(\textbf{B})= 0$.

Once it leaves the magnetic field, the particle travels freely
until it reaches the screen, $P_1$, placed a distance $D=20~cm$
beyond the magnet.

\subsection{The Probability Density in the Magnetic Field}

The particle stays within the magnetic field for a time $\Delta
t=\frac{\Delta l}{v}= 2\times 10^{-5} s$. During this time
$[0,\Delta t]$, the spinor is (see Appendix A)
\begin{equation}\label{eq:fonctiondanschampmagnétique}
\Psi (x,z,t)= \left(
\begin{array}{c}
                                R_{+}e^{i\frac{S_{+}}{\hbar }} \\
                                R_{-}e^{i\frac{S_{-}}{\hbar }}
                            \end{array}
                     \right)
\simeq \left(
\begin{array}{c}
                                \cos \frac{\theta_0}{2}
                 (2\pi\sigma_0^2)^{-\frac{1}{2}}
                 e^{-\frac{(z-\frac{\mu_{B} B'_{0}}{2 m}t^{2})^2 + x^2}
                 {4\sigma_0^2}} e^{i\frac{\mu_{B} B'_{0}t z -\frac{\mu^2_{0} B'^2_{0}}{6 m}t^3 +  \mu_B B_0 t + \frac{\hbar \varphi_0}{2}}{\hbar }}\\
                                i \sin \frac{\theta_0}{2}
                 (2\pi\sigma_0^2)^{-\frac{1}{2}}
                 e^{-\frac{(z+\frac{\mu_{B} B'_{0}}{2 m}t^{2})^2 + x^2}
                 {4\sigma_0^2}} e^{i\frac{-\mu_B B'_{0}t z -\frac{\mu^2_{0} B'^2_{0}}{6
    m}t^3 -  \mu_B B_0 t -\frac{\hbar \varphi_0}{2}}{\hbar }}
                            \end{array}
                     \right)
\end{equation}

Since the initial spinor direction is random, the atomic density,
$\rho(z,t)$ is found by integrating $R_+^2+R_{-}^2$ on $(\theta_0,
\varphi_0)$ and $x$ (notice that $R_+^2+R_{-}^2$ is independent of
$\varphi_0$). So one gets:
\begin{eqnarray}\label{eq:densitédanschampmagnétique}
    \rho(z,t) &=
     (2\pi\sigma_0^2)^{-\frac{1}{2}}
                  \frac{1}{2}\left(e^{-\frac{(z-\frac{\mu_{B} B'_{0}}{2 m}t^{2})^2}{2\sigma_0^2}}+
                  e^{-\frac{(z+\frac{\mu_{B} B'_{0}}{2
                  m}t^{2})^2}{2\sigma_0^2}}\right).
\end{eqnarray}

\subsection{The Probability Density after the Magnetic Field}

After the magnetic field, at time $t+ \Delta t$ $(t \geq 0)$, the
spinor becomes (see Appendix A)
\begin{equation}\label{eq:fonctionapreschampmagnétique}
\Psi (x,z,t+\Delta t)= \left(
\begin{array}{c}
                                R_{+}e^{i\frac{S_{+}}{\hbar }} \\
                                R_{-}e^{i\frac{S_{-}}{\hbar }}
                            \end{array}
                     \right)
\simeq \left(
\begin{array}{c}
                                \cos \frac{\theta_0}{2}
                 (2\pi\sigma_0^2)^{-\frac{1}{2}}
                 e^{-\frac{(z-z_{\Delta}- ut)^2 + x^2}
                 {4\sigma_0^2}} e^{i\frac{m u z + \hbar \varphi_+}{\hbar }} \\
                                i \sin \frac{\theta_0}{2}
                (2\pi\sigma_0^2)^{-\frac{1}{2}}
                e^{-\frac{(z+z_{\Delta}+ ut)^2 + x^2}
                {4\sigma_0^2}} e^{i\frac{-
    muz + \hbar \varphi_-}{\hbar }}
                            \end{array}
                     \right)
\end{equation}

 where

\begin{equation}\label{eq:zdeltavitesse}
    z_{\Delta}=\frac{\mu_B B'_{0}(\Delta
    t)^{2}}{2 m}=10^{-5}m,~~~~~~u =\frac{\mu_B B'_{0}(\Delta t)}{m}=1 m/s.
\end{equation}

One can deduce (as previously) the atom density $\rho$ at (z,$t+
\Delta t$):
\begin{eqnarray}\label{eq:densitéaprèschampmagnétique}
    \rho(z,t+ \Delta t) &=
     (2\pi\sigma_0^2)^{-\frac{1}{2}}
                  \frac{1}{2}\left(e^{-\frac{(z-z_{\Delta}- ut)^2}{2\sigma_0^2}}+
                  e^{-\frac{(z+z_{\Delta}+
                  ut)^2}{2\sigma_0^2}}\right).
\end{eqnarray}

\begin{figure}[H]
\begin{center}
\includegraphics[width=.2\linewidth]{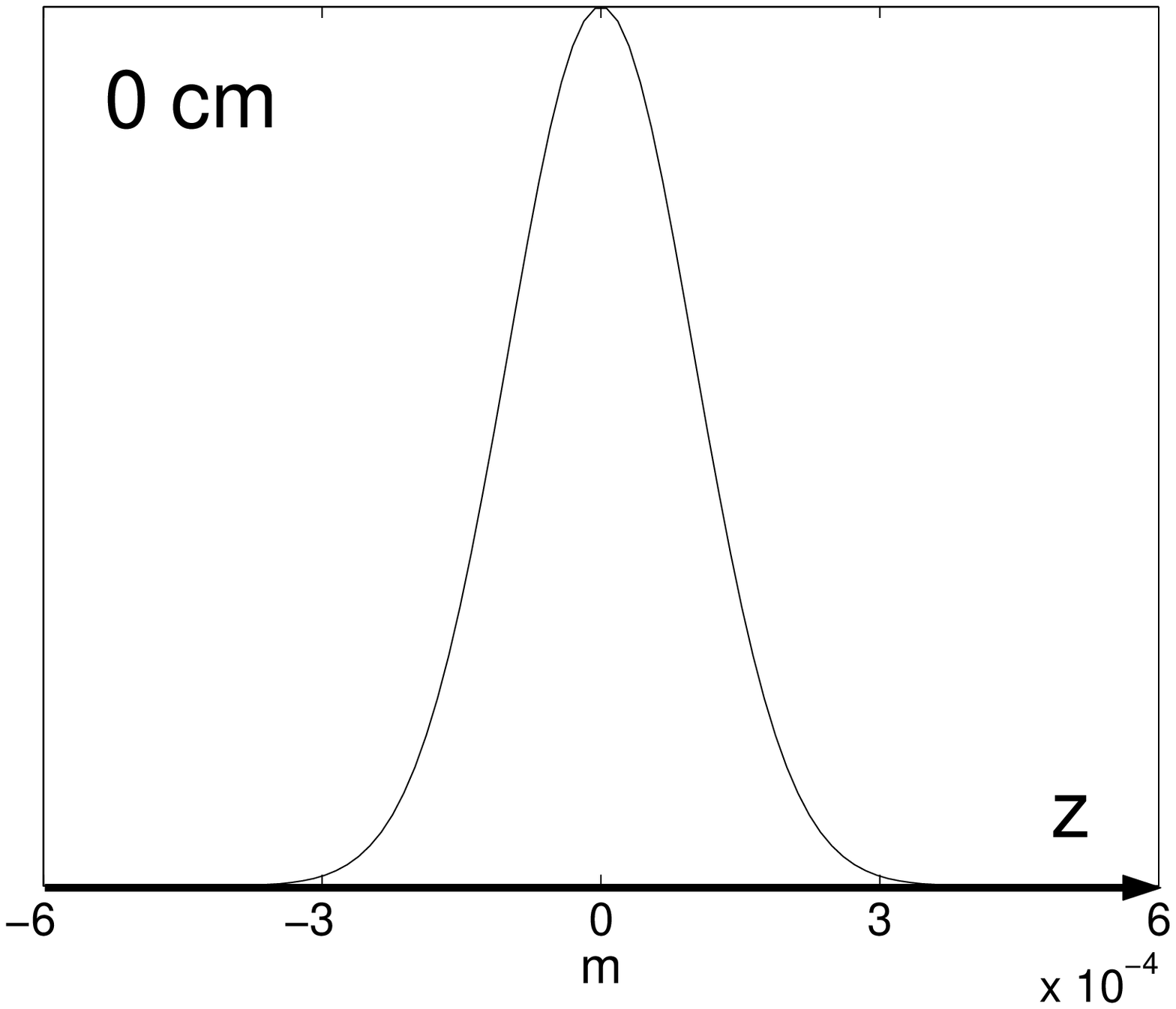}
\includegraphics[width=.2\linewidth]{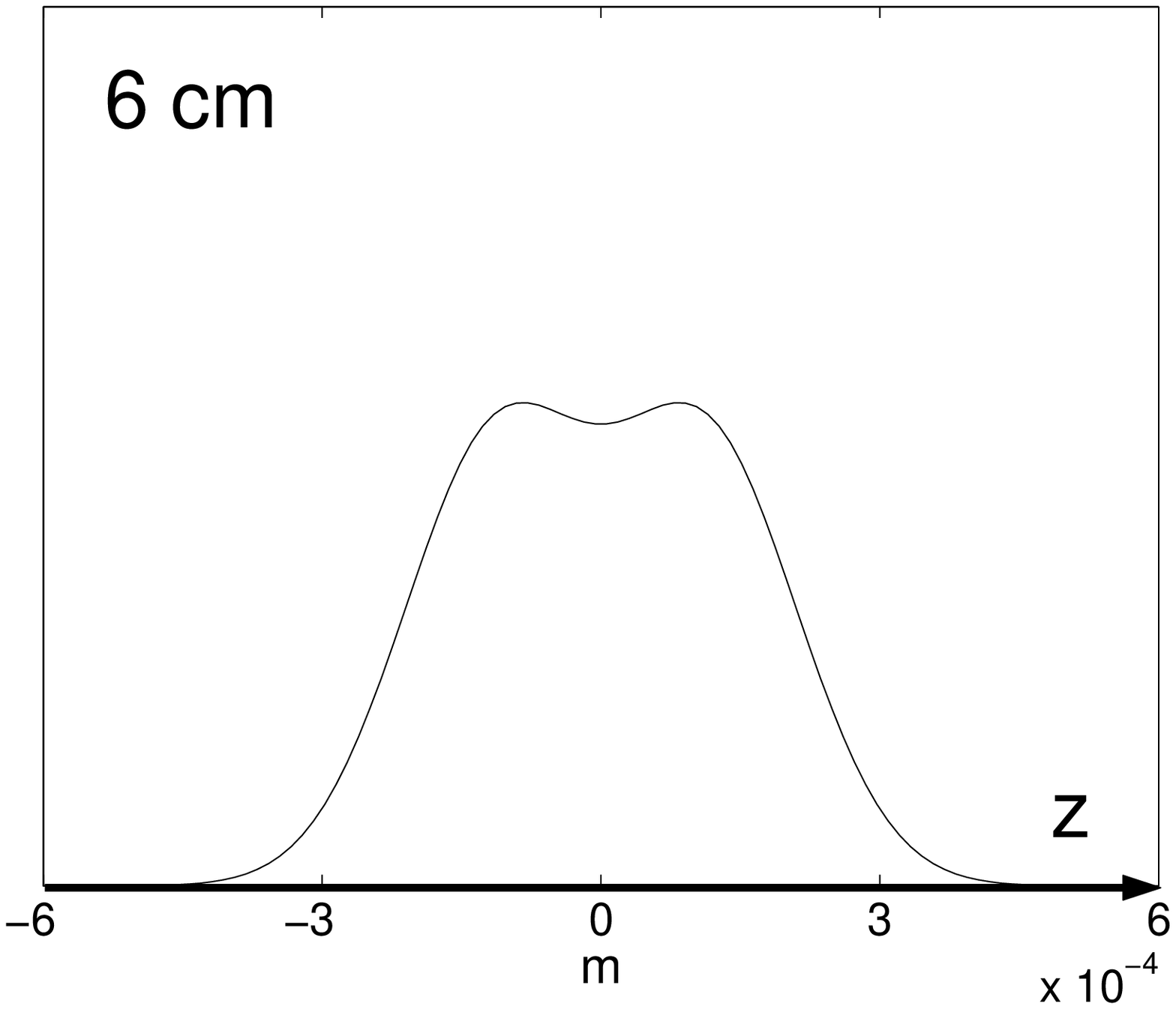}
\includegraphics[width=.2\linewidth]{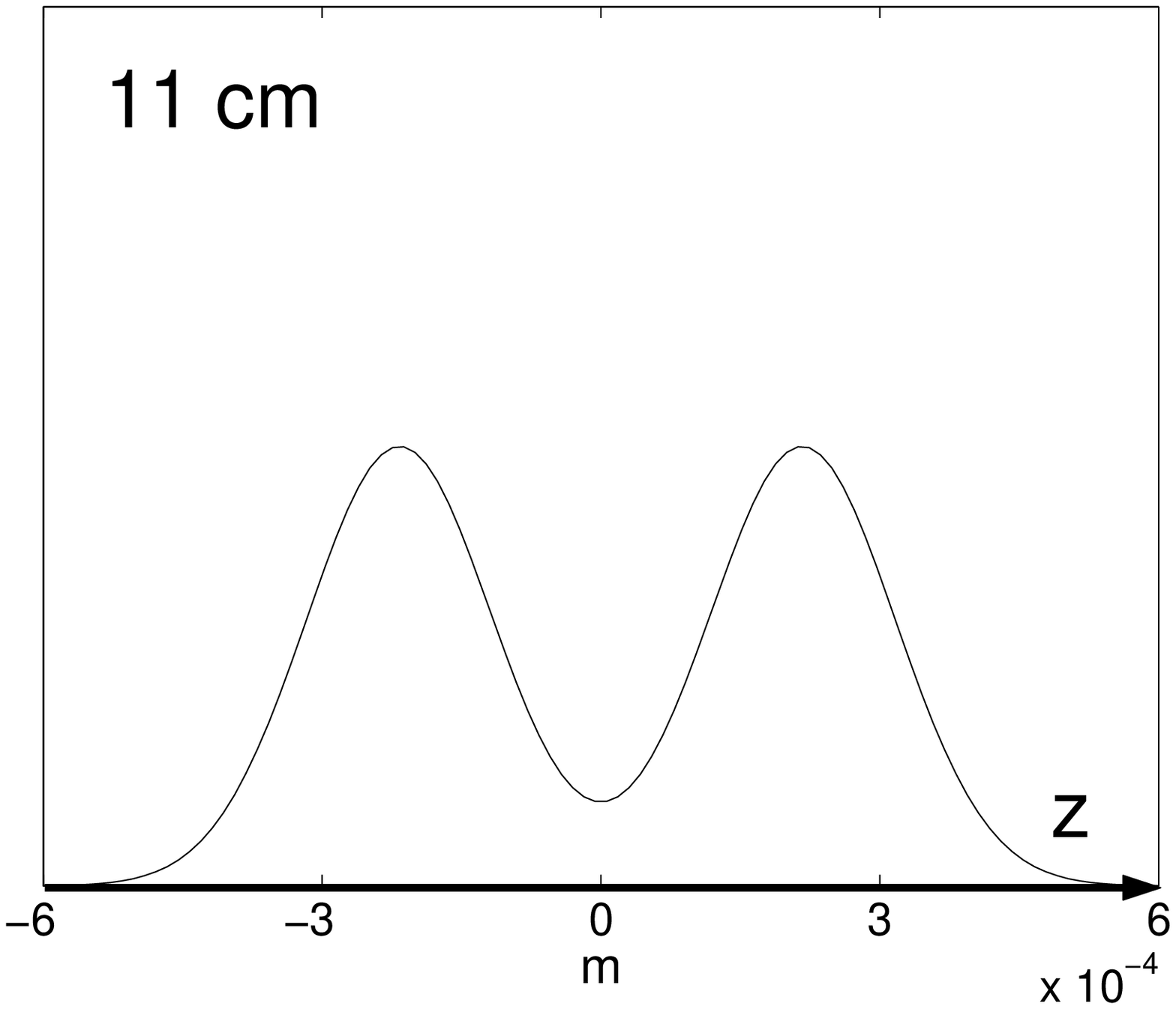}
\includegraphics[width=.2\linewidth]{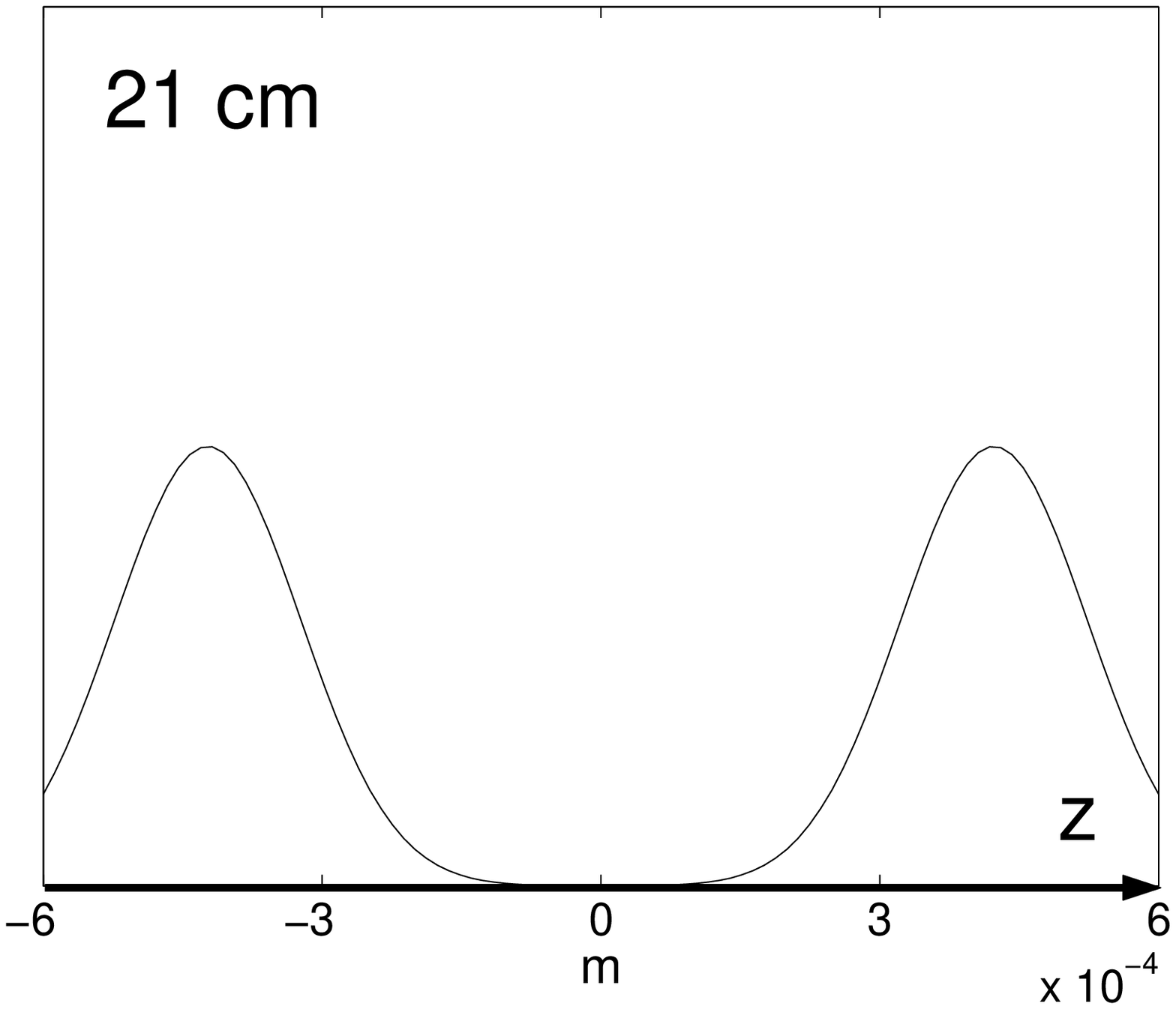}
\caption{\label{fig:ddp-SetG}Evolution of the probability density
of silver atoms.}
\end{center}
\end{figure}

Figure ~\ref{fig:ddp-SetG} shows the probability density of the
silver atoms as a function of z at several values of t ( the plots
are labelled with $y = vt$). The beam separation does not appear
at the end of the magnetic field (1 cm), but 10 cm further along.

In equations (\ref{eq:densitédanschampmagnétique}) and
(\ref{eq:densitéaprèschampmagnétique}) one recognizes classical
trajectories:
\begin{equation}\label{eq:trajectoiresclassiques}
z_{\pm}=\{\pm \frac{\mu_B B'_{0} t^{2}}{m} ~~for~~ t\in[0, \Delta
t]; \pm z_{\Delta} \pm u t ~~for~~ \Delta t+t (t\geq0)\}
\end{equation}
corresponding to particles with magnetic moments of
$\mu_{z}=\mu_{B}$ and $\mu_{z}=-\mu_{B}$ respectively. The
trajectories are parabolic inside the magnetic field and linear
after. The two spots $N^{+}$ and $N^{-}$ become separated when
$z_{+} - z_{-} \gtrapprox 4 \sigma_{0}$, which occurs at the
separation time
\begin{equation}\label{eq:tempsdecoherence}
t_{s} \simeq \frac{3 \sigma_{0}}{u}= \frac{3 \sigma_{0} m v}{\mu_B
B'_{0} \Delta l}=\frac{6 \sigma_{0} m m_e v}{e \hbar B'_{0} \Delta
l}=3 \times 10^{-4}s.
\end{equation}

Statistically, everything happens as if the atom's moment were
quantified in two parts: one half of the particles having
$\mu_{z}=\mu_{B}$ and the other half having $\mu_{z}=-\mu_{B}$.
This result explains why the semi-classical description of this
Stern-Gerlach experiment is usually used. This semi-classical
description starts from the quantization of spin and deduces from
Ehrenfest's Theorem the average trajectories of the spinors with
initial spinors $\Psi_{+}^{0}=(2\pi\sigma_{0}^{2})^{-\frac{1}{2}}
                      e^{-\frac{(z^2+x^2)}{4\sigma_0^2}}
                      \binom{1}
                            {0}$ and $\Psi_{-}^{0}=(2\pi\sigma_{0}^{2})^{-\frac{1}{2}}
                      e^{-\frac{(z^2+x^2)}{4\sigma_0^2}}
                      \binom{0}
                            {1}$ respectively.
Note that this is a statistical interpretation, and the
individuality of the atoms represented by the angles $\theta_{0}$
and $\varphi_{0}$ is lost. However experimentally, one does not
observe directly the wave function, but \textit{individual}
impacts of the silver atoms on the $P_{1}$ screen. The usual
explanation of these individual impacts on the screen is
decoherence, \cite{Zurek,Schlosshauer} caused by the interaction
with the measurement device. Here, the evolution of the
probability density given by equations
(\ref{eq:densitédanschampmagnétique}) and
 (\ref{eq:densitéaprèschampmagnétique}) correctly describes a separation of the wave
packet into two packets thanks to the measurement device, but
cannot describe the individual positions of these impacts.



\section{Impacts and Quantization explained by trajectories}

To explain individual impacts, we simulate the silver atom
trajectories in the de Broglie-Bohm interpretation just as we did
\cite{Gondran_2005} for the neon atoms in the Young's double slit
experiment. In this unusual presentation of the Quantum Mechanics
results, the particle is represented not only by its wave
function, but also by the position of its center-of-mass.

Indeed at the first instant, the wave function $\Psi^{0}(x,z)$
gives the initial probability density $\Psi^{0}(x,z)$. This
density, which doesn't depend on $\hbar$, is the classic presence
density of a silver atom. So in classic mechanics, one does have
an undetermination on the atom position, and in order to describe
its evolution, it is necessary to precise its initial position.
The principle of the De Broglie-Bohm interpretation is to do the
same in quantum mechanics.

So, in the Pauli equation case (\ref{eq:Pauli}), atoms have
trajectories that are defined by using the center-of-mass velocity
$\textbf{v}(x,y,z,t)$ given\cite{Bohm_1955,Takabayasi_1954} by :
\begin{equation}\label{eq:vitesseGordon}
     \textbf{v} = \frac{\hbar}{2m \rho} Im{(\Psi^\dag\boldsymbol\nabla \Psi)}
                + \frac{\hbar }{2 m \rho}
     rot(\Psi^\dag\boldsymbol\sigma\Psi)
\end{equation}
where $\Psi^\dag=[\psi_{+}^{*}, \psi_{-}^{*}]$.

Let us show how this interpretation gives the same results as the
Copenhagen school. One verifies\cite{Bohm_1955,Takabayasi_1954}
that with $\textbf{v}$ given by (\ref{eq:vitesseGordon}), the
probability density $\rho(x,y,z,t)=\Psi^\dag
\Psi=|\Psi(x,y,z,t)|^{2}$ of the $\Psi$ spinor solution of the
Pauli equation (\ref{eq:Pauli}) satisfies the Madelung continuity
equation:
\begin{equation}\nonumber
    \frac{\partial \rho }{\partial t}+div(\rho \textbf{v})=0.
\end{equation}
One can deduce from it, that if a particle family with the initial
probability density $\rho_{0}(x,y,z)$ follows the de Broglie-Bohm
trajectories, its probability density at the t time, will be
$\rho(x,y,z,t)$.

Thus, these two interpretations give statistically identical
results, but the de Broglie-Bohm interpretation predicts the
position of individual impacts as well. We shall see that these
trajectories also explain the spin quantization following the
magnetic field gradient.

In equation (\ref{eq:vitesseGordon}), the last term $\frac{\hbar
}{2 m \rho}
     rot(\Psi^\dag\boldsymbol\sigma\Psi)$ corresponds to
the Gordon current. Its contribution to velocity is here
negligible. We will therefore not take it into account from now.

In the de Broglie-Bohm interpretation, the individual particle is
not only described by its wave function, but by its initial
position $(x_{0},z_{0},y_{0}=0)$ as well. So, trajectories in x
and z are given by the differential equations:
\begin{equation}\label{eq:vitessex}
     \frac{d x}{d t}=\frac{1}{2
     m}\frac{\partial(S_{+}+S_{-})}{\partial x}+\frac{1}{2 m}\frac{\partial(S_{+}-S_{-})}{\partial x}\cos\theta
\end{equation}
\begin{equation}\label{eq:vitessez}
     \frac{d z}{d t}=\frac{1}{2
     m}\frac{\partial(S_{+}+S_{-})}{\partial z}+\frac{1}{2 m}\frac{\partial(S_{+}-S_{-})}{\partial z}\cos\theta
\end{equation}
with $\tan\frac{\theta}{2} = \frac{R_{-}}{R_{+}}$.

A silver atom with a polarization ($\theta_0$,$\varphi_0$) and a
position $z_0$ at the entrance of the electro-magnet $A_1$ will
satisfy the differential equation in the $[0,\Delta t]$ period:
\begin{equation}\label{eq:trajectoiredanschampSetG}
 \frac{d z}{d t}=\frac{\mu_B B'_{0} t}{ m} cos\theta(z,t) ~~~~with ~~~~~~\tan \frac{\theta(z,t)}{2}=
 \tan\frac{\theta_0}{2} e^{- \frac{\mu_B B'_{0} t^{2}z}
{2 m \sigma_0^{2}}}
\end{equation}
and for the $\Delta t +t$ ($t\geq0$) period:
\begin{equation}\label{eq:trajectoireapreschampSetG}
 \frac{d z}{d t}=u \frac{\tanh(\frac{(z_\Delta + ut) z}{\sigma_{0}^{2}})+
 \cos \theta_0}{1+\tanh(\frac{(z_\Delta + ut) z}{\sigma_{0}^{2}})\cos
 \theta_0}~~~~and ~~~~~~\tan \frac{\theta(z,t)}{2}=\tan\frac{\theta_0}{2} e^{- \frac{(z_\Delta + ut)z}{\sigma_0^{2}}}.
\end{equation}

\begin{figure}[H]
\begin{center}
\includegraphics[width=0.6\linewidth]{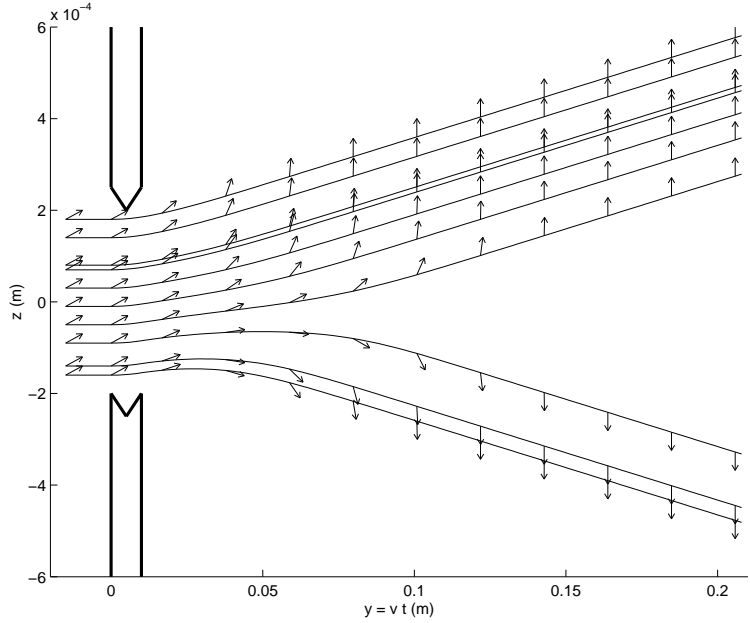}
\caption{\label{fig:SetG-10traj} Ten silver atom trajectories with
initial polarization $(\theta_0=\frac{\pi}{3})$ and initial
position $z_0$; Arrows represent the spin orientation
$\theta(z,t)$ along the trajectories.}
\end{center}
\end{figure}

Figure~\ref{fig:SetG-10traj} presents a plot in the x,y plane of a
set of 10 silver atom trajectories with the initial polarization
$(\theta_0=\frac{\pi}{3},\varphi_0=0)$ and whose initial position
$z_0$ has been randomly chosen from a Gaussian distribution with
standard deviation $\sigma_{0}$. The spin orientations
$\theta(z,t)$ are represented by arrows.

Figure~\ref{fig:SetG-10traj} presents, for a silver atom with the
initial polarization $(\theta_0=\frac{\pi}{3},\varphi_0=0)$, a
plot in the x,y plane of a set of 10 trajectories whose initial
position $z_0$ has been randomly chosen from a Gaussian
distribution with standard deviation $\sigma_{0}$. The spin
orientations $\theta(z,t)$ are represented by arrows.

One can notice that the final orientation, obtained after the
separation time $t_{s}$, will depend on the initial particle
position $z_{0}$ in the wave packet and on the initial angle
$\theta_{0}$ of the atom magnetic moment with the axis z.

We obtain $+\frac{\pi}{2}$ if $z_0
> z^{\theta_{0}}$ and $-\frac{\pi}{2}$ if $z_0
< z^{\theta_{0}}$ with
\begin{equation}\label{eq:seuilpolarization}
z^{\theta_{0}}=\sigma_0 F^{-1}(\sin^{2}\frac{\theta_{0}}{2})
\end{equation}
where F is the cumulative distribution function of the standard
normal distribution.

So besides explaining the position of impacts, this simulation
shows that it is possible to give a simple interpretation of
quantization on z-axis.

\begin{figure}[H]
\begin{center}
\includegraphics[width=0.6\linewidth]{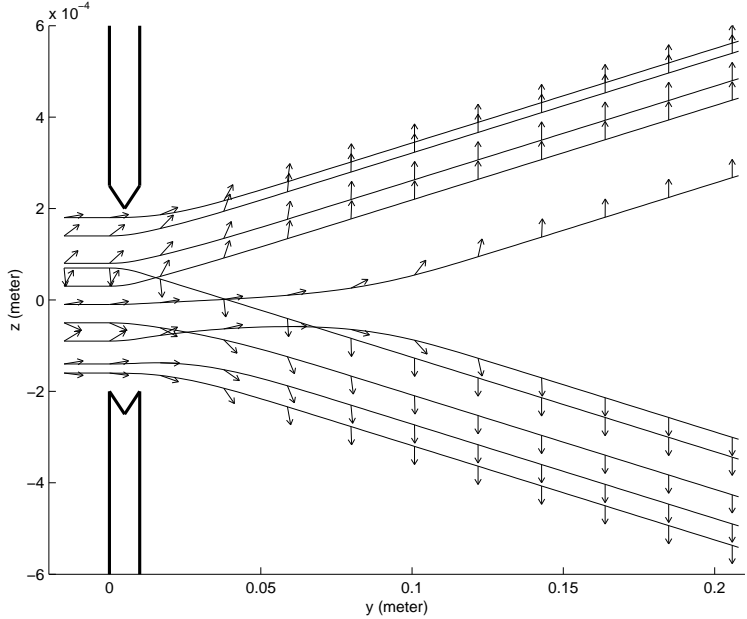}
\caption{\label{fig:SetG-10trajectoires}10 silver atom
trajectories after the electro-magnet; Arrows represent the spin
orientation $\theta(z,t)$ along the trajectories.}
\end{center}
\end{figure}

Figure~\ref{fig:SetG-10trajectoires} presents a plot in the x,y
plane of a set of 10 silver atom trajectories whose initial
characteristics $(\theta_0,\varphi_0,z_0)$ have been randomly
chosen: $\theta_0$ and $\varphi_0$ from an uniform distribution
and $z_0$ from a Gaussian distribution with standard deviation
$\sigma_{0}$. The spin orientations $\theta(z,t)$ are represented
by arrows.

\section{Conclusion}

It is now possible to propose the following interpretation of
measurement in quantum mechanics. There certainly exists an
interaction with the measuring apparatus as it is classically
explained, and there exists a minimum delay necessary for the
measurement $t_s=\frac{3 \sigma_{0}}{u}$. Yet we believe that this
measurement and this delay do not have the \textit{meanings} that
are classically attributed to them. In the present case, the
measuring apparatus itself gives the orientation of the spin
either in the direction of field, or in the direction opposite to
the field, depending on the position of the particle in the wave
packet. The measuring delay is then the time which is necessary
for the particle to point its spin to the final direction.

Let us notice that in this numerical study of the Stern-Gerlach
experiment we didn't need any of the classical postulates of
measurement in quantum mechanics, such as eigenvalues of the
Hamiltonian or wave packet reduction. These two postulates may
account for experimental results, but they do not give any idea of
the transitions that lead to these results. Instead we only used
the quantum mechanics equations (Pauli equation) and made only one
hypothesis : the centers of mass of the atoms in the atomic beam
are spacially distributed according to the density given by the
wave function, and follow paths that are compatible with the
continuity equation (De Broglie-Bohm hypothesis). From this one
and only hypothesis, we provided altogether:

- a simple explanation of the position of the particle impacts;

- a simple explanation of the spin quantization along the
measurement axis ;

- an simple explanation of the transition towards the Hamiltonian
eigenvalues.

\bigskip
\appendix

\section{Calculating the spinor of the Stern-Gerlach experiment}

In the magnetic field $B=(B_x,0,B_z)$, the Pauli equation
(\ref{eq:Pauli}) gives coupled Schrödinger equations for each
spinor component
\begin{equation}\label{eq:Paulicomplet}
     i\hbar \frac{\partial\psi_{\pm}}{\partial t}(x,z,t)= - \frac{\hbar^2}{2 m} \nabla^{2} \psi_{\pm }(x,z,t)  \pm
     \mu_B (B_0 - B_0' z) \psi_{\pm}(x,z,t) \mp i \mu_B B_0' x
     \psi_{\mp}(x,z,t).
\end{equation}
If one effects the transformation \cite{Platt}
\begin{equation}\nonumber
     \psi_{\pm}(x,z,t)= \exp (\pm \frac{i \mu_B B_0 t}{\hbar})
     \overline{\psi}_{\pm}(x,z,t)
\end{equation}
equation~(\ref{eq:Paulicomplet}) becomes
\begin{equation}\nonumber
     i\hbar \frac{\partial\overline{\psi}_{\pm}}{\partial t}(x,z,t)= -\frac{\hbar^{2}}{2 m} \nabla^{2} \overline{\psi}_{\pm}(x,z,t)
     \mp \mu_B
     B'_0 z \overline{\psi}_{\pm}(x,z,t) \mp i \mu_B
     B'_0 x \overline{\psi}_{\mp}(x,z,t)\exp(\pm i \frac{2 \mu_B B_0 t}{\hbar})
\end{equation}
The coupling term oscillates rapidly with frequency $\omega=
\frac{2 \mu_B B_0}{\hbar}=1,4 \times 10^{11} s^{-1}$. Since
$|B_{0}|\gg |B'_{0}z|$ and $|B_{0}|\gg |B'_{0}x|$, the period of
oscillation is short compared to the motion of the packet along
its trajectory. Avering over a period that is long compared to the
period of oscillation, causes the coupling term to vanish,
yielding \cite{Platt}
\begin{equation}\nonumber
     i\hbar \frac{\partial\overline{\psi}_{\pm}}{\partial t}(x,z,t)= -\frac{\hbar^{2}}{2 m} \nabla^{2} \overline{\psi}_{\pm}(x,z,t)
     \mp \mu_B
     B'_0 z \overline{\psi}_{\pm}(x,z,t).
\end{equation}

The initial wave function
$\overline{\psi}_{\pm}^{0}(x,z)=\psi_{\pm}^{0}(x,z)=\psi_{x}^{0}(x)
\psi_{\pm}^{0}(z)$ with
$\psi_{x}^{0}(x)=(2\pi\sigma_{0}^{2})^{-\frac{1}{4}}
                      e^{-\frac{x^2}{4\sigma_0^2}}$, $\psi_{+}^{0}(z)=(2\pi\sigma_{0}^{2})^{-\frac{1}{4}}
                      e^{-\frac{z^2}{4\sigma_0^2}}
                      \cos \frac{\theta_0}{2}e^{ i\frac{\varphi_0}{2}}$ and $\psi_{-}^{0}(z)=(2\pi\sigma_{0}^{2})^{-\frac{1}{2}}
                      e^{-\frac{z^2}{4\sigma_0^2}}i\sin\frac{\theta_0}{2}e^{-i\frac{\varphi_0}{2}}$ allows a separation of variables x and z.
Then we can compute explicitly the preceding equations for all t
in $[0, \Delta t]$ with $\Delta t=\frac{\Delta l}{v}=2 \times
10^{5}s$.

We obtain: $\overline{\psi}_{\pm}(x,z,t)=\overline{\psi}_{x}(x,t)
\overline{\psi}_{\pm}(z,t)$ with
\begin{equation}\label{eq:ondelibregauss}
\overline{\psi}_{x}(x,t)=(2\pi\sigma_{t}^{2})^{-\frac{1}{4}}
                      e^{-\frac{x^2}{4\sigma_t^2}}\exp \frac{i}{\hbar}[-\frac{\hbar}{2}\tan^{-1}(\frac{\hbar t}{2 m \sigma_{0}^{2}})
                      +\frac{x^{2}\hbar^{2}t^{2}}{8 m \sigma_{0}^{2} \sigma_{t}^{2}}]
\end{equation}

\begin{equation}\nonumber
\overline{\psi}_{+}(z,t)= \psi_{K}(z,t) \cos \frac{\theta_0}{2}e^{
i\frac{\varphi_0}{2}}~~~~~and~~K=-\mu_B B'_{0}
\end{equation}

\begin{equation}\nonumber
\overline{\psi}_{-}(z,t)= \psi_{K}(z,t)
i\sin\frac{\theta_0}{2}e^{-i\frac{\varphi_0}{2}}~~~~~and~~K=+\mu_B
B'_{0}
\end{equation}
$\sigma_t^2 = \sigma_0^2 + \left(\frac{\hbar
t}{2m\sigma_0}\right)^2$ and
\begin{equation}\label{eq:ondegauss}
 \psi_{K}(z,t)=(2\pi\sigma_{t}^{2})^{-\frac{1}{4}}
                      e^{-\frac{(z+\frac{K t^{2}}{2 m})^2}{4\sigma_t^2}}\exp \frac{i}{\hbar}
                      [-\frac{\hbar}{2}\tan^{-1}(\frac{\hbar t}{2 m \sigma_{0}^{2}})
                      -Ktz-\frac{K^{2}t^{3}}{6 m}
                      +\frac{(z+\frac{K t^{2}}{2 m})^{2}\hbar^{2}t^{2}}{8 m \sigma_{0}^{2}
                      \sigma_{t}^{2}}].
\end{equation}
Equations~(\ref{eq:ondelibregauss}) and~(\ref{eq:ondegauss}) are
classical results.\cite{Feynman}

The experimental conditions give $\frac{\hbar \Delta t}{2 m
\sigma_0}=4 \times 10^{-11}~m \ll \sigma_{0}=10^{-4}~m$. We deduce
the approximations $\sigma_{t} \simeq  \sigma_0$,
$\overline{\psi}_{x}(x,t) \simeq
\psi_{x}^{0}(x)=(2\pi\sigma_{0}^{2})^{-\frac{1}{4}}
                      e^{-\frac{x^2}{4\sigma_0^2}}$ and
\begin{equation}\label{eq:ondegaussapprox}
 \overline{\psi}_{K}(z,t)  \simeq (2\pi\sigma_{0}^{2})^{-\frac{1}{4}}
                      e^{-\frac{(z+\frac{K t^{2}}{2 m})^2}{4\sigma_0^2}}\exp \frac{i}{\hbar}
                      [ -Ktz-\frac{K^{2}t^{3}}{6 m}].
\end{equation}

At the end of the magnetic field, at time $\Delta t$, the spinor
equals to
\begin{equation}\label{eq:ondeadeltat}
    \Psi (x,z,\Delta t) = \psi_{x}(x,\Delta t) \left( \begin{array}{c}
                                \psi_{+}(z,\Delta t)\\
                                \psi_{-}(z,\Delta t)
                            \end{array}
                     \right)
\end{equation}
with
\begin{equation}\nonumber
\psi_{+}(z,\Delta t)=(2\pi\sigma_{0}^{2})^{-\frac{1}{4}}
                      e^{-\frac{(z - z_{\Delta})^2}{4\sigma_0^2}+ \frac{i}{\hbar}m u z}  \cos \frac{\theta_0}{2}e^{
i \varphi_{+}}
\end{equation}
\begin{equation}\nonumber
\psi_{-}(z,\Delta t)= (2\pi\sigma_{0}^{2})^{-\frac{1}{4}}
                      e^{-\frac{(z + z_{\Delta})^2}{4\sigma_0^2}- \frac{i}{\hbar}m u z} i\sin\frac{\theta_0}{2}e^{i \varphi_{-}}
\end{equation}
\begin{equation}\nonumber
    z_{\Delta}=\frac{\mu_B B'_{0}(\Delta
    t)^{2}}{2 m},~~~~~~u =\frac{\mu_0 B'_{0}(\Delta t)}{m}~~~~and
\end{equation}
\begin{equation}\nonumber
    \varphi_{+}=\frac{\varphi_{0}}{2} -\frac{\mu_{B}B_0 \Delta
    t}{\hbar}-\frac{K^{2}(\Delta t)^{3}}{6 m \hbar};~~~~\varphi_{-}=-\frac{\varphi_{0}}{2} +\frac{\mu_{0}B_0 \Delta
    t}{\hbar}-\frac{K^{2}(\Delta t)^{3}}{6 m \hbar}.
\end{equation}

We remark that the crossing through the magnetic field gives the
equivalent of a velocity $+u$ in the direction $0z$ to the
function $\psi_+$ and a velocity $-u$ to the function $\psi_{-}$.
Then we have a free particle with the initial wave
function~(\ref{eq:ondeadeltat}). The Pauli equation resolution
gives again $\psi_{\pm}(x,z,t)=\psi_{x}(x,t) \psi_{\pm}(z,t)$ and
with the experimental conditions we have $\psi_{x}(x,t) \simeq
(2\pi\sigma_{0}^{2})^{-\frac{1}{4}} e^{-\frac{x^2}{4\sigma_0^2}}$
and
\begin{equation}\nonumber
\psi_{+}(z,t +\Delta t) \simeq (2\pi\sigma_{0}^{2})^{-\frac{1}{4}}
                      e^{-\frac{(z - z_{\Delta}- u t)^2}{4\sigma_0^2}+ \frac{i}{\hbar}(m u z -\frac{1}{2}m u^2 t + \hbar \varphi_{+})}
                        \cos \frac{\theta_0}{2}
\end{equation}
\begin{equation}\nonumber
\psi_{-}(z,t + \Delta t) \simeq
(2\pi\sigma_{0}^{2})^{-\frac{1}{4}}
                      e^{-\frac{(z + z_{\Delta}+u t)^2 }{4\sigma_0^2}+\frac{i}{\hbar}(- m u z -\frac{1}{2}m u^2 t + \hbar \varphi_{-})}
                       i\sin\frac{\theta_0}{2}
\end{equation}

\end{document}